\documentclass{iaus}
\usepackage{graphicx}
\newcommand{\apj}{Astrophys. J.}
\newcommand{\aj}{Astron. J.}
\newcommand{\aap}{Astr. Ap.}
\newcommand{\apjl}{Astrophys. J. (Letters)}
\newcommand{\apjs}{Astrophys. J. Suppl.}
\newcommand{\pasp}{Pub. Astr. Soc. Pacific}
\newcommand{\mnras}{Mon. Not. Roy. Astr. Soc.}
\newcommand{\nat}{Nature}
\newcommand{\araa}{Ann. Rev. Astr. Astroph.}
\newcommand{\lsun}{\mbox{L$_\odot$}}% Lsun
\newcommand{\msun}{\mbox{M$_\odot$}}% Msun
\newcommand{\rsun}{\mbox{R$_\odot$}}% Rsun
\newcommand{\lbol}{\mbox{$L_{bol}$}} % bolometric luminosity
\newcommand{\tbol}{\mbox{$T_{bol}$}} % bolometric temperature
\newcommand{\ee}[1]{\mbox{${} \times 10^{#1}$}}% scientific number format
\newcommand\cmv{\mbox{cm$^{-3}$}}

\newcommand{\msunmyr}{\mbox{M$_\odot$ Myr$^{-1}$}}% Msun per million years
\newcommand{\tk}{\mbox{$T_K$}}
\newcommand{\kms}{\mbox{km s$^{-1}$}}% km/s
\newcommand{\mstar}{\mbox{$M_{\star}$}}
\newcommand{\msunyr}{\mbox{M$_\odot$ yr$^{-1}$}}% Msun per year

\title[JD 11.~~Low-mass Star Formation: Observations]
{Low-mass Star Formation: Observations}

\author[Neal J. Evans II]   %% give here short author list %%
{Neal J. Evans II$^1$}
\affiliation{$^1$Department of Astronomy, The University of Texas at Austin,
1 University Station, C1400 Austin, TX 78712-0259, USA
\\ email: {\tt nje@astro.as.utexas.edu} }

\pubyear{2010}
\volume{xxx}  %% insert here IAU Symposium No.
\pagerange{119--126}
\jname{Computational Star Formation}
\editors{A.C. Editor, B.D. Editor \& C.E. Editor, eds.}
\begin{document}

\maketitle

\begin{abstract}
I briefly review recent observations of regions forming low mass stars.
The discussion is cast in the form of seven questions that have 
been partially answered, or at least illuminated, by new data. 
These are the following:
where do stars form in molecular clouds; what determines the IMF; how long
do the steps of the process take; how efficient is star formation; do any
theories explain the data; how are the star and disk built over time; and
what chemical changes accompany star and planet formation. I close with
a summary and list of open questions.
%% add here a maximum of 10 keywords, to be taken form the file <Keywords.txt>
\end{abstract}

\firstsection % if your document starts with a section,
              % remove some space above using this command.
\section{Introduction}\label{sec1}

Recent large-scale surveys of nearby molecular clouds have provided
a solid statistical basis for addressing some long-standing questions
about the formation of low-mass stars. Ideally, one would like unbiassed
surveys at wavelengths ranging from radio to X-ray of all molecular clouds
within some radius of the Sun. Radio and X-ray surveys probe non-thermal
and transient events; millimeter continuum surveys probe the mass and structure
of the dust, thereby tracing the gas, the far-infrared contains the luminosity
information for embedded stages, the mid-infrared probes the disk, the
near-infrared probes the inner disk and star, and the visible and ultraviolet
probe the star and ongoing accretion. Together, these define the most
basic properties of the molecular cloud material and the forming star.
While we will concentrate on the continuum in this review, spectroscopic
information provides vital complementary information (e.g., see \S \ref{sec8}). 
Again, a wide wavelength range for spectroscopy is ideal.

The most significant recent additions to our arsenal include large scale
surveys of the nearby clouds in the Gould Belt in the mid-infrared with
Spitzer (\cite[Evans et al. 2003]{evans03}, \cite[Evans et al. 2009]{evans09}, 
\cite[Allen et al. 2010]{allen10} and references therein) 
and in the millimeter continuum 
(\cite[Enoch et al. 2006]{enoch06}, \cite[Young et al. 2006]{young06}, 
\cite[Enoch et al. 2007]{enoch07}, \cite[Enoch et al. 2008]{enoch08}, 
\cite[Enoch et al. 2009]{enoch09}). 
The Spitzer-based surveys will soon by joined by a complete survey with 
Herschel spanning the wavelengths between mid-infrared and millimeter 
(\cite[Andr{\'e} et al. 2010]{andre10} and Andr{\'e} in this volume). 
Sometime later,
the same clouds will be surveyed completely in millimeter continuum with
SCUBA-2 on the JCMT (\cite[Ward-Thompson et al. 2007]{ward-thompson07}).

\section{Where do Stars Form in Molecular Clouds?}\label{sec2}

Spitzer surveys of 20 clouds in the Gould Belt, now including nearly
all clouds within 500 pc, have provided a much more complete data base
of Young Stellar Objects (YSOs) (\cite[Evans et al. 2009]{evans09}, \cite[G{\"u}del et al. 2007]{guedel07}, 
\cite[Rebull et al. 2010]{rebull10}, \cite[Allen et al. 2010]{allen10}). 
These surveys have clearly established
that star formation is not evenly distributed over molecular clouds.
Instead, star formation proceeds in a clustered manner, concentrated in
regions of high extinction (\cite[J{\o}rgensen et al. 2008]{joergensen08}, 
\cite[Evans et al. 2009]{evans09}).
Using the criterion of \cite{lada03} of 1 \msun\ pc$^{-3}$, 91\%
of the stars in the c2d survey (\cite[Evans et al. 2009]{evans09}) and 75\% of those
in a larger sample, including Taurus and all Gould Belt clouds, but
excluding Orion, form in a clustered environment 
(\cite[Bressert et al. 2010]{bressert10}).
Clustering is a matter of degree, with a broad range of surface densities
and no clear sign of separate ``clustered" and ``distributed" modes.
Furthermore, the degree of clustering is stronger in younger SED classes,
such as Class I and pre-stellar cores, suggesting that the Class II sources
have dispersed slightly since their formation.

Using only the Class I and Flat SED classes to avoid any dispersal issues,
\cite[Heiderman et al. (2010)]{heiderman10} have found evidence for a threshold gas surface density
above which star formation is much faster. This threshold lies roughly at
$A_V = 8$ mag, or 120 \msun\ pc$^{-2}$. A very similar threshold was
identified by \cite{lada10} from an independent analysis. Since most
of the mass of molecular clouds lies below this threshold, at least in
local clouds, a threshold may yield insight into why star formation
efficiencies are so low in molecular clouds (see \S \ref{sec5}).

\section{What Determines the IMF?}\label{sec3}

This age-old question has been revivified by catalogs of prestellar
cores that show a mass function somewhat similar to that of stars
but shifted to higher masses (e.g., \cite[Motte et al. 1998]{motte98}, 
\cite[Johnstone et al. 2000]{johnstone00}, 
\cite[Enoch et al. 2008]{enoch08}, \cite[Sadavoy et al. 2010]{Sadavoy10}).  
Similar distributions were found for starless,
but unbound and hence not prestellar (see \cite[di Francesco et al. 2007]{difrancesco07} for
definitions) cores in the Pipe Nebula (\cite[Lada et al. 2008]{lada08}). 
These results
are consistent with a picture in which the IMF is set by the process
of core formation and favor a picture in which about 1/3 of the core
mass is incorporated into the star. However, arguments have been
advanced that some cores will evolve faster [either the less massive
(\cite[Clark et al. 2007]{clark07}) or the more massive (\cite[Hatchell \& Fuller 2008]{hatchell08})],
and hence the current core mass function
is not consistent with a one-one mapping onto the IMF. \cite{swift08}
have explored other issues, and
\cite{reid10} have argued that existing measurements of the clump
mass spectrum are highly compromised by limitations of the observational
techniques, such as sensitivity, spatial resolution, and spatial filtering
of large scale emission. While these
theoretical issues will be hard to resolve, the statistics should at
least improve as the Herschel Gould Belt surveys become available
(for a preview, see article by Andr{\'e} in this volume).

\section{How Long Do the Stages of the Process Take?}\label{sec4}

The major evolutionary organizing principle for star formation has
been the Class system, first introduced by \cite{lada87} and extended by
\cite{greene94} and \cite{andre93}, using the shape of the SED, measured
in various ways to establish a putative evolutionary scheme from
prestellar cores to Class 0 to Class I to Flat SED to Class II to Class III. 
Recently, \cite{robitaille06} has usefully distinguished the SED Class
from the physical Stages, which proceed from Prestellar to Stage 0 (most
of mass still in envelope) to Stage I (most of mass in central star and disk
but still substantial envelope) to Stage II (envelope gone, but substantial
disk) to Stage III (disk mostly gone, mass in star).
The history of this system has been reviewed by \cite{evans09}.
The connection of Class to Stage is generally accepted, but orientation
effects can easily confuse a neat identification of Class with Stage:
for example, a face-on Stage 0 source may be classified as a Class I source,
while an edge-on Stage I source could have a Class 0 SED.

The method of assigning durations to these Classes is to count the numbers
in each Class, relating the numbers to the durations by pinning all durations
to one that is assumed to be known. One has to further assume that 
star formation has been continuous over a time longer than the duration of
the class used to calibrate ages and that other variables, such as mass,
are not important. With these caveats in mind, the results of the c2d
studies of nearby clouds with Spitzer led to longer durations for Class I
and Class 0 than had previously been derived. The durations are all pinned
to a timescale of 2 Myr for the Class II phase, based on studies of
infrared excess frequency in clusters of known ages. That age is uncertain
by $\pm 1$ Myr and should be thought of as half-life, rather than a
fixed duration for each individual star. Using the latest, but still
preliminary, statistics
from the combined c2d plus Gould Belt projects (3124 YSOs in 20 clouds),
the Class I duration would be $0.55\pm0.28$ Myr and the Flat SED lasts 
$0.36\pm0.18$ Myr, where the uncertainties are due to the uncertainty in the
Class II duration.
In this calculation, Class 0 is included with Class I. If they are
separated using the \tbol\ criterion, the Class 0 phase lasts 0.16 Myr
(\cite[Enoch et al. 2009]{enoch09}, \cite[Evans et al. 2009]{evans09}), substantially longer than earlier estimates.
Using the millimeter surveys together with the infrared to separate out
the prestellar cores, \cite{enoch08} found a duration of 0.43 Myr for
prestellar cores once
the mean density of the core exceeded the detection threshold of about
$n = 2\ee4$ \cmv. Most cores had higher mean densities, and more careful
comparison suggested a duration roughly 3 times the free-fall time
(\cite[Enoch et al. 2008]{enoch08}). This is another area that should be revolutionized by
Herschel observations of the Gould Belt clouds.

\section{How Efficient is Star Formation?}\label{sec5}

The c2d and Gould Belt surveys provide a uniform analysis of a much
more complete sample of YSOs. Eventually the Taurus cloud and some other
clouds should be brought into this analysis. Currently, we use the
same 3124 YSOs ($N_{YSOs}$) in 20 clouds discussed in \S \ref{sec4}. 
These surveys are about 90\% complete to about
0.05 \lsun. The total mass of YSOs is calculated from $M_* = N_{YSOs} \times
0.5$ \msun, where 0.5 \msun\ is the mean stellar mass. The star formation
rate is then $SFR = M_*/t_{II}$ \msunmyr, where $t_{II}$ is the duration of the 
Class II phase in Myr. The YSO counts are incomplete for Class III objects,
so we are effectively computing a star formation rate averaged over the 
last 2 Myr. For efficiencies, we compare the mass in stars to the mass
in the molecular cloud, measured from extinction mapping over the same
region in which YSOs are counted (generally $A_V \ge\ 2$ mag). For clouds
with millimeter continuum maps, the mass in dense gas can be computed 
from the sum over masses of all the dense cores.

Based on the sum of all stellar and cloud masses for all 20 c2d and Gould
Belt clouds, 
about 3\% of the mass is in YSOs younger than 2 Myr and the mass in dense
cores ranges from 2\% to 5\% for the three clouds with complete maps. In
contrast, the total mass in YSOs is very similar to that in dense cores,
suggesting that star formation is rapid and efficient once the dense cores
have formed. Clearly, star formation is not very efficient for the clouds
as a whole, but the efficiency of star formation is quite high once dense
cores have formed.

While not efficient in an absolute sense, star formation in the local
clouds is much more efficient than would be predicted from the 
Kennicutt-Schmidt relations used in extragalactic studies. The relations
predict the star formation rate surface density from the gas surface density
(e.g., \cite[Kennicutt 1998]{kennicutt98}). While a few of the least active clouds in the
Gould Belt sample lie on the extragalactic relation, almost all lie well above,
most by an order of magnitude (\cite[Evans et al. 2009]{evans09}). This discrepancy has
been explored in detail by \cite{heiderman10}, who find evidence for
a threshold surface density, as discussed above. They suspect that
the measures of gas surface density in other galaxies, based almost entirely
on CO emission, include large amounts of gas below the threshold 
(\S \ref{sec2}).

\section{Do Any Theories Explain the Data?}\label{sec6}

The existing data provide some reality checks for star formation theories.
For example, the most commonly used picture of low-mass star formation
is inside-out collapse (\cite[Shu 1977]{shu77}). In this picture, collapse begins at 
the center of a singular isothermal sphere, and a wave of infall propagates
outward at the sound speed ($c_s$). The end of infall is less defined in this
model, but it is usually assumed that, when the infall wave reaches the outer
boundary of the core, the remainder of the core falls in. The time for this
is equal to the time taken for the infall wave to reach the outer boundary.
(For a more correct treatment of the evolution of a core with a distinct
boundary, see \cite[Vorobyov \& Basu 2005a]{vorobyov05a}).
For a kinetic temperature, $\tk = 10$ K, $c_s = 0.19$ \kms.
If we associate the time indicated above for the Class I phase of 0.55 Myr 
with the final infall of the last bit of envelope, the wave of infall would
have reached the outer boundary in 0.55/2 Myr. The corresponding radius would
be $r_{out} = c_s t_I/2 = 0.055$ pc, roughly consistent with many of the 
core size distributions (e.g., \cite[Enoch et al. 2008]{enoch08}) and the mean
separation of YSOs in clusters of $0.072\pm0.006$ pc (\cite[Gutermuth et. al 2009]{gutermuth09}).

The density distribution of the initial state 
is set by the temperature in the simplest version of this model. Thus,
the mass available in the mean duration of a Class I source is also set.
If a fraction $f$ ends up in the star,
the resulting stellar mass is $\mstar = 0.86 f$ \msun. If $f = 0.3$,
as suggested above, the resulting mass is 0.26 \msun, near the mode of the
IMF. Thus a simple picture of cores undergoing inside-out collapse meets
some general consistency checks.

However, serious problems appear when we compare the luminosity function
of the YSOs in the c2d sample to predictions for the Shu model. The
Shu model predicts a mass infall rate of $\dot M_{inf} = 1.6\ee{-6}$ \msunyr.
The radiation released by this infall onto an object of 0.08 \msun,
at the boundary between brown dwarfs and stars, and radius of 3 \rsun\
produces a luminosity $L_{acc} = 1.6$ \lsun. Most (59\%) of the YSOs in the
c2d clouds lie below this luminosity (\cite[Dunham et al. 2010]{dunham10}). More generally,
the distribution of sources in the \tbol-\lbol\ plane (\cite[Evans et al. 2009]{evans09}
is very poorly
represented by tracks that follow the Shu solution and use radiative
transfer to calculate the SED as a function of time (\cite[Young \& Evans 2005]{young05}).

These issues are not peculiar to the Shu model. In fact, almost all other
models feature faster infall and will produce a still larger discrepancy
with the data. One is faced with the problem of decreasing the typical
(observed) accretion rate onto the star, while still removing the
envelope in about 0.5 Myr and building the star.

\section{How Are the Star and Disk Built over Time?}\label{sec7}

The problem of low luminosities and a very large spread (at least
three orders of magnitude) in luminosity must be telling us about
the process of growth of the disk and star. Because the vast majority
of the gravitational energy is released when matter accretes onto the star,
rather than onto the disk, storing matter in the disk is an obvious option.
Since there is no reason that the accretion rate from disk to star should
be synchronized with the infall rate onto the disk, it is very plausible
that material accumulates in the disk until an instability causes a fast
matter transfer to the star, followed by a slower rebuilding of the disk
from the envelope. This picture of episodic accretion explains the high
incidence of low luminosity values (little or no accretion onto the star) 
and the large spread in luminosity (small numbers of sources caught 
in a high-accretion state).

Indeed, \cite{kenyon90} suggested this solution to the problem of low
luminosity, which was apparent even with IRAS data. The Spitzer data has
only exacerbated the problem, and the episodic accretion solution is 
even more attractive.
Furthermore, simulations of the flow from envelope to disk to star show
instabilities and rapid variations in accretion onto the star 
(\cite[Vorobyov \& Basu 2005b]{vorobyov05b}, \cite[Vorobyov \& Basu 2006]{vorobyov06}). Observationally, there is direct
evidence in the form of FU Orionis events (\cite[Hartmann \& Kenyon 1996]{hartmann96}), outflow
morphologies showing multiple ejection events (e.g., \cite[Lee et al. 2007]{lee07}), and
studies of low luminosity sources with extended outflows. These last studies
show that minimum mean luminosities needed to drive the observed outflows
exceed the current luminosity of the YSO (\cite[Dunham et al. 2006]{dunham06}, 
\cite[Dunham et al. 2010a]{dunham10a}).

To put these ideas on firmer footing, \cite[Dunham et al. 2010b]{dunham10b} 
explored a simple
model of episodic accretion, including full 2D radiative transfer with
outflows removing matter from a \cite{terebey84} rotating collapse model.
He could reproduce reasonably well the observed luminosity distribution
and the distribution in the \tbol-\lbol\ plane.

If accretion onto the star is episodic, there are substantial consequences.
First, the connection between Stages and Classes becomes even more tenuous, as
the luminosity flares can move an object in one Stage back and forth across
Class boundaries. Second, the luminosity is not an indicator of stellar mass
until nuclear burning dominates accretion luminosity. Third, there may be 
long-term consequences for the star, causing incorrect estimates of stellar 
ages (\cite[Baraffe et al. 2009]{baraffe09}) and possibly the low lithium
abundances in young stars (\cite[Baraffe et al. 2010]{baraffe10}). 
Finally, the initial conditions for planet formation
may be determined by the timing of the last accretion event: if it happens
just before the last of the envelope falls onto the disk, the star will
start with a low-mass disk; conversely, if the disk was close to its maximum
stable mass when the last of the envelope accretes, the star would have
a massive disk.

\section{What Chemical Changes Accompany Star and Disk Formation?}\label{sec8}

While we have focused on the macroscopic changes as material flows from
envelope to disk to star, microscopic changes to the chemical state
occur during the process. Infrared spectroscopy from the ground and
from Spitzer have revealed a rich spectrum of ices on dust grains
before (\cite[Knez et al. 2005]{knez05}) and during (\cite[Boogert et al. 2008]{boogert08}, \cite[Pontoppidan et al. 2008]{pontoppidan08},
\cite[{\"O}berg et al. 2008]{oberg08}, \cite[Bottinelli et al. 2010]{bottinelli10}) the infall phase. 
The size distribution of dust
grains shifts to larger grains in molecular clouds, as compared to the
diffuse interstellar medium (\cite[Flaherty et al. 2007]{flaherty07}, \cite[Chapman et al. 2009]{chapman09}). 
A new model of dust opacities in dense cores that incorporates these
aspects, especially the ice features, is needed to fit SED data that
include mid-infrared spectroscopy (e.g., \cite[Kim et al. 2010]{kim10}).

The flip side of the formation of ices is severe depletion of the gas
phase species via freeze-out onto grains (for reviews, see
\cite[Ceccarelli et al. 2007]{ceccarelli07}, \cite[van Dishoeck 2009]{ewine09}). This can reach extreme levels
in prestellar cores. As a forming star increases in luminosity, central
heating can evaporate some ices, producing a complex pattern of
abundance (\cite[J{\o}rgensen et al. 2004]{joergensen04}). To interpret molecular line observations
correctly, one should ideally use fully self-consistent chemo-dynamical
models (\cite[Lee et al. 2004]{lee04}, \cite[Evans et al. 2005]{evans05},
\cite[Chen et al. 2009]{chen09}) 
that track the chemistry during
collapse. Episodic accretion models further complicate the picture, as
the changes in luminosity can subject the grains to the equivalent of
freeze-thaw cycles that may drive greater differentiation and complexity.
For a concrete example, CO frozen on grains may be converted to CO$_2$
(\cite[Pontoppidan et al. 2008]{pontoppidan08}, which does not evaporate as easily as the CO;
repeated cycles can systematically deplete the CO in the envelope
(\cite[Kim et al. 2010]{kim10}).

When the envelope material falls onto the disk, some ices may evaporate
(\cite[Watson et al. 2007]{watson07}) and outflows will drive further thermal
and chemical changes along the outflow
walls, which are nicely traced by very high-J CO lines observable with
Herschel PACS spectroscopic observations (\cite[van Kempen st al. 2010a]
{kempen10a}, \cite[van Kempen st al. 2010b]{kempen10b}).

\section{Summary and Open Questions}\label{sec9}

The main points are summarized here.

\begin{enumerate}
\item{Stars form mostly in clustered environments, but the degree of clustering
varies smoothly over a wide range.}
\item{The core mass function is similar to the initial mass function of stars,
but shifted to higher masses. If the core mass function maps to the IMF,
the shift suggest an efficiency of about 1/3. However, there are many 
caveats in this discussion.}
\item{The timescales for the Class 0 and Class I SED classes are longer
than previously thought. Based on number counts, they are roughly 0.16 Myr
for Class 0 and 0.55 Myr for Class I.}
\item{Efficiencies for star formation over the last 2 Myr are low
($\sim 3$\%) in nearby clouds, but quite high ($>25$\%) in dense cores.}
\item{Star formation rate surface densities are much higher for a given
gas surface density that would be predicted by Kennicutt-Schmidt relations.}
\item{Both the discrepancy with the Kennicutt-Schmidt relations and
the low absolute efficiency are likely related to a surface density 
threshold for efficient star formation that is not met by the great
majority of the cloud mass in nearby clouds, nor in normal galaxies.}
\item{A simple inside-out collapse picture for a core is consistent with
much of the data on sizes and masses, but no model with uniform accretion
reproduces the distribution of YSOs in the \tbol-\lbol\ plane.
Episodic accretion is strongly indicated.}
\item{Complex chemical changes accompany the large-scale flows of matter,
as material moves from gas to ice and back. Simple models of abundances that
are constant, either in time or space, can be misleading.}
\end{enumerate}

The most significant open questions are summarized here.
\begin{itemize}
\item{What sets the mass of stars? Is it the core mass function, or is
it feedback?}
\item{How do brown dwarfs form? A core forming a brown dwarf would have
to be very small and very dense to be bound.}
\item{What controls the threshold for star formation? While a surface
density threshold is indicated by the data, this may be a secondary
indicator for a more complex set of variables, including volume density,
turbulent velocity, magnetic field, ionization, etc.}
\item{And finally a question that has been with us since the early 
days of molecular clouds: why is star formation so inefficient?}
\end{itemize}

I would like to acknowledge the collaboration of A. Heiderman, M. Dunham,
L. Allen, D. Padgett, and E. Bressert in preparing this review. 
Support for this work, part of the \textit{Spitzer} Legacy Science Program, 
was provided by NASA through contracts 1224608 and 1288664 issued by the 
Jet Propulsion Laboratory, California Institute of Technology, 
under NASA contract 1407. Support was also
provided by NASA Origins grant NNX07AJ72G and NSF grant AST-0607793
to the University of Texas at Austin.

%\begin{discussion}

%\end{discussion}

\end{document}